\documentclass[iop,apj]{emulateapj}
\usepackage{apjfonts}

\usepackage{amsmath}
\usepackage{psfig}
\usepackage{bm}

\newcommand{\unitstyle}[1]{\ensuremath{\mathrm{#1}}}

\newcommand{\Edot}{\ensuremath{\dot{E}}} 

\newcommand{\Mdot}{\ensuremath{\dot{M}}} 

\newcommand{\Msun}{\ensuremath{\unitstyle{M}_\odot}}

\begin{document}
\title{Mechanical feedback from black hole accretion as an energy source of core-collapse supernova explosions}

\author{En-hao Feng, Rong-feng Shen, Wei-peng Lin}
\affiliation{School of Physics and Astronomy, Sun Yat-sen University, Zhuhai, China}


\email{Corresponding author: R-FS}
\email{E-HF, fengenh@mail2.sysu.edu.cn}
\email{R-FS, shenrf3@mail.sysu.edu.cn}

\begin{abstract}
According to the traditional scenario for core-collapse supernovae, the core of the collapsing star forms a neutron star and its gravitational energy release sends out a shockwave into the stellar envelope. However, in a significant number of numerical simulations, the shock stalls and the star cannot be exploded successfully, especially for a massive, compact star. We consider an alternative scenario that with mass fallback, the collapsing star forms a black hole in the center, surrounded by a dense, hot accretion disk, which blows out an intense outflow (wind). The kinetic energy of the wind may result in a successful stellar explosion. With an improved version of the formulism in Kohri et al. (2005) who studied neutron star accretion of minor fallback, we study this disk wind-driven explosion by calculating the accretion history for a suite of pre-SN stellar models with different initial surface rotational velocities, masses and metallicities, and by comparing the disk wind energy with the binding energy of the infalling stellar envelope.
We show that the most promising models to be exploded successfully by this new channel are those relatively compact pre-SN stars with relatively low metallicities and not too low specific angular momenta. The total energies of the explosions are $\sim 10^{51-52}$ergs, and a more massive progenitor may produce a more energetic explosion.

\end{abstract}


\section{Introduction}
Core-collapse supernovae (CCSNe) are a high energy phenomenon caused by the collapse of a massive star. In the traditional core collapse supernova theory, the core of a massive star collapses to form a neutron star (NS). When the subsequent falling material reach the surface of the neutron star, there will be a rebound shock. If the explosion energy is large enough, the shock wave is able to propagate to the surface of the progenitor, and most of the material will be unbound. On the other hand, if the explosion energy is not large enough, then the shock wave will stall at a radius deep inside the collapsing star. In order to cause a successful supernova explosion in this case, the revival of the shock is needed.

When the core collapses to the proto-NS, there is an intense neutrino emission about $10^{53}$erg, which may help for the shock revival under some conditions (O'Connor \& Ott 2011; Ugliano et al. 2012; Pejcha \& Thompson 2015). Other mechanisms like the standing accretion shock instability (e.g., Blondin et al. 2003; Blondin \& Mezzacappa 2007; Hanke et al. 2013), and the collapse-induced thermonuclear explosion (e.g., Kushnir \& Katz 2015) are also suggested to explain CCSNe. However, if the revival is unsuccessful, then most of the material will fall back, rendering a `failed' explosion. This situation has happened in a considerable number of numerical simulations (Herant et al. 1994; Rampp \& Janka 2000; Liebend\"orfer et al. 2001; Thompson et al. 2003; Sumiyoshi et al. 2005; Woosley \& Janka
2005; Janka et al. 2007; Janka 2012; Hanke et al. 2013; Dolence et al. 2015; Melson et al. 2015a; Skinner et al. 2016; Suwa et al. 2016; Janka et al. 2016). Therefore, new channels of exploding stars are called for.

The fallback of enough mass makes the further collapse of the NS into a black hole (BH) inevitable.
About 20 years ago, the collapsar model, in which the core of the massive star collapses to a BH surrounded by a compact disk, whose accretion generates a relativistic jet, was developed to explain the long gamma-ray bursts (GRBs; Woosley 1993; Popham et al. 1999; MacFadyen \& Woosley 1999; MacFadyen et al. 2001; Woosley \& Heger 2012). Later, type Ic SNe (e.g., Herger et al. 2003) and long GRBs were found to be associated (e.g., Galama et al. 1998; Stanek et al. 2003). Woosley \& Bloom (2006) gave a review for SNe-GRBs connection. Also, GRBs and type Ic SNe are found to occur in similar regions of their host galaxies (Kelly et al. 2008).
It is therefore not inconceivable that the collapsar model is also responsible for the supernovae associated with GRBs (MacFadyen 2003). Milosavljevi\'c et al. (2012) studied the detailed shock propagation in GRBs-associated SNe, and with simulations (Lindner et al. 2012). Also, some GRBs such as long-duration GRBs are suggest to be explained by failed SNe with fallback accretion (Fynbo et al. 2006; Fryer et al. 2007; Quataert \& Kasen 2012).

We take a step further, and postulate that many CCSNe are driven by BH accretion, but not limited to those associated with GRBs. We consider a new channel to explode the star. That is, the explosion is powered by a disk wind. Figure 1 is a cartoon of this model.

\begin{figure*}
\begin{minipage}[t]{0.5\textwidth}
\centering
\includegraphics[width=18cm]{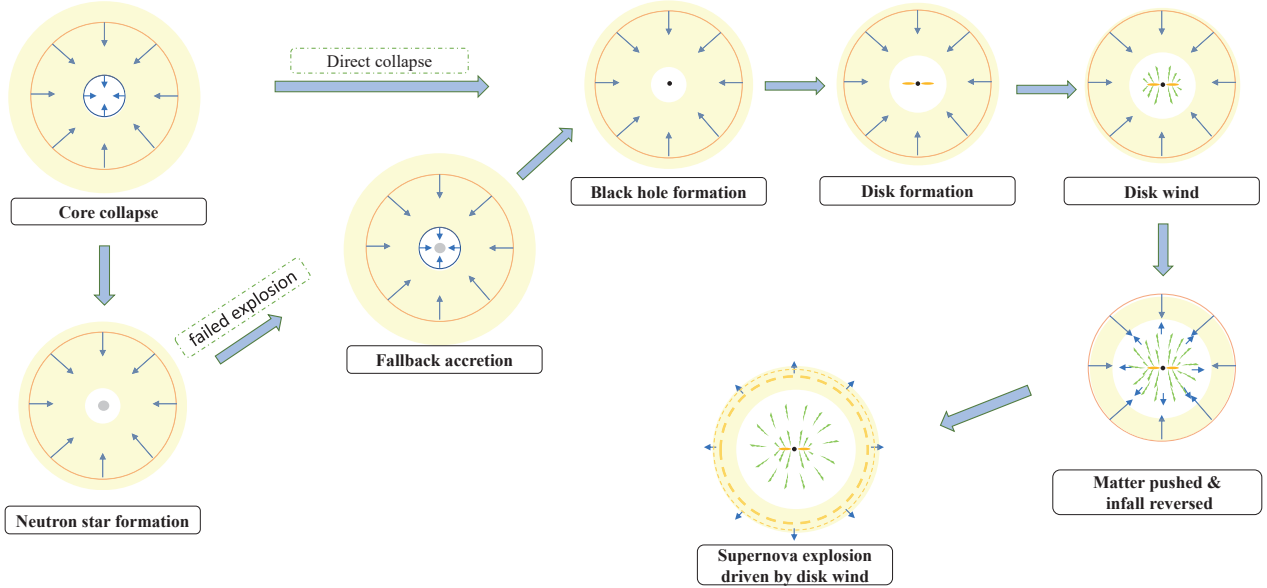}
\end{minipage}
\caption{A cartoon about the SN driven by disk wind. The core of the massive star either collapses to a BH directly, or collapses to a neutron star firstly, then with fallback accretion collapses to a BH. For the second case, however, following the neutron star formation, if the initial explosion energy is large enough, there may be no material remaining bound to the neutron star, but we will not discuss this situation here. After the BH formation, an accretion disk forms if the specific angular momentum of the subsequent falling matter is large enough. A vigorous disk wind is expected to launch if the accretion flow is advection dominated. As the wind power rises sharply, the falling matter cannot resist the kinetic energy of the wind and will be pushed away, resulting in a successful explosion.}
\label{fig:Illustrating}
\end{figure*}

The GRBs-associated SNe are all type I's. For Type II SNe, the standard neutrino-driven explosion mechanism (Janka 2012) is feasible if the required explosion energy is not much larger than $10^{51}$erg. However, some Type II SNe are found with explosion energies be much larger than $10^{51}$erg (Botticella et al. 2010; Utrobin et al. 2010; Moriya et al. 2013). Other mechanism such as fallback accretion may be possible to explain these high energy Type II SNe. Moriya et al. (2018) used fallback accretion mechanism to explain the light curves of Type II SN OGLE-2014-SN-073, whose explosion energy is about $10^{52}$erg. Also, the light curves of super-luminous supernovae (SLSNe) may be explained by fallback accretion (Dexter \& Kasen 2013).

Because we consider black hole accretion, we consider the cases with BH remnants. O'Connor \& Ott (2011) studied the conditions of zero-age main sequence (ZAMS) that leave a BH remnants in failing CCSNe.
For the progenitors that leave BH remnant, there may be two scenarios after core collapse process. The first is that the core collapses into a black hole directly. This can happen when the initial mass of the star is large. The second case is that a neutron star forms first, and it further collapses into a BH with fallback accretion. Whether the fallback matter will form a disk or not is determined by the distribution of specific angular momentum of the star at the pre-SN state. The latter is in turn determined by the initial rotation of the ZAMS star, subject to changes during its lifetime due to mass loss and internal mixing (which inturn depend on star's mass and metallicity).
Perna et al. (2014) used some MESA (Paxton et al. 2011; 2013; 2015; 2018) -generated pre-SN models to discuss the disk formation and evolution, and we will use these pre-SN models to assess the possibilities that they will be exploded by black hole accretion.

Due to those stellar parameter dependence, there is a wide range of outcome for BH accretion. Kashiyama \& Quataert (2015) discussed the case that most of the core and envelope directly falls into the BH, only the outermost layer has the specific angular momentum large enough to form a disk. This lightweight explosion will produce a fast luminous blue transient. This will happen if the progenitors are blue supergiants (BSGs) or Wolf-Rayet stars (WRs). For red supergiants (RSGs), even a weak initial explosion can unbind a few $M_\odot$ of envelope (Lovegrove \& Woosley 2013). Nevertheless, Horiuchi et al. (2014) studied the failed SNe in the RSG case.

When the accretion rate is extremely high, the cooling of the disk is dominated by neutrino emission and the disk is a neutrino-dominated accretion flow (NDAF; Popham et al. 1999; Narayan et al. 2001; Di Matteo et al. 2002; Kohri \& Mineshige 2002; Janiuk et al. 2004, 2007; Lee \& Ramirez-Ruiz 2006; Surman et al. 2006; Chen \& Beloborodov 2007; Shibata et al. 2007). At lower accretion rates, the disk cools very inefficiently and becomes an advection-dominated accretion flow (ADAF; Narayan \& Yi 1994, 1995; Lee, Ramirez-Ruiz \& Page 2004, 2005; Kohri et al. 2005; Kumar et al. 2008), for which a strong mass outflow is expected (Narayan \& Yi 1994, 1995; Stone et al. 1999; Igumenshchev \& Abramowicz 2000; Narayan et al. 2000; Quataert \& Gruzinov 2000; Narayan, Piran \& Kumar 2001).

Kohri et al. (2005) calculated the energy output of an outflow (disk wind) from an ADAF around a central NS of 1.4 $M_{\odot}$. Here, we extend the calculation to the BH accretion disk in the collapsar scenario, and include the time dependence. Moreover, we tie the history of accretion (thus, of the wind energy output) to the pre-SN property of the star. The latter in turn tracks back to the ZAMS property of the star via MESA calculation (Perna et al. 2014).

The paper is structured as follows. In Section 2, we describe the disk wind mechanism, and calculate the wind energy.
In Section 3, we introduce the fallback accretion and disk formation process.
In Section 4, we apply the disk wind model to some pre-SN models provided by Perna et al. (2014). These models are organized in a grid of 2 metallicities, 3 masses and 3 initial angular velocities.
In section 5, we present our results and compare the wind energy with the gravitational binding energy of the still infalling envelope to determine the feasibility of a successful wind-driven explosion and the mass of the ejecta.
We summarize our findings in section 6 and discuss the implications in Section 7.

\section{The disk wind energy}

It has long been found that, at highly super-Eddington mass supply rate, an accretion disk is in the so-called ADAF regime, which tends to launch an intense outflow (disk wind; Narayan \& Yi 1994, 1995; Stone et al. 1999; Igumenshchev \& Abramowicz 2000; Narayan et al. 2000; Quataert \& Gruzinov 2000; Narayan, Piran \& Kumar 2001; Kohri et al. 2005).
For even higher accretion rates, or for the inner region of the disk, the accretion flow is a NDAF (Popham et al. 1999; Narayan et al. 2001; Di Matteo et al. 2002; Kohri \& Mineshige 2002; Janiuk et al. 2004, 2007; Lee \& Ramirez-Ruiz 2006; Surman et al. 2006; Chen \& Beloborodov 2007; Shibata et al. 2007) and it cools efficiently via neutrino emission.
\begin{figure}
\centering
\includegraphics[width=8cm]{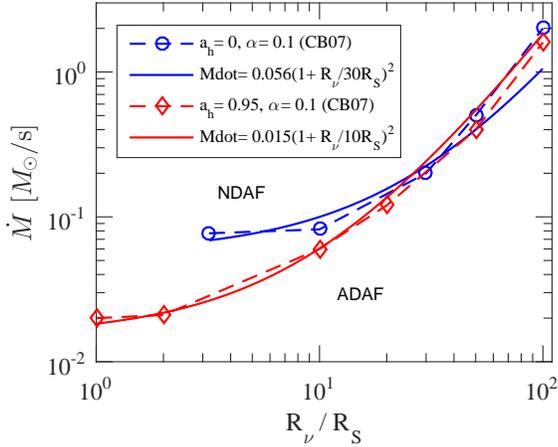}
\caption{The relation between the NDAF / ADAF transition radius $R_{\nu}$ and the local accretion rate $\dot M (R_{\nu})$. The data points (marked as CB07 in the legends) are adapted from Figure 11 of Chen \& Beloborodov (2007) which was calculated for a 3$M_{\odot}$ black hole. The solid lines are empirical functions (i.e., Equation (\ref{eq:MdotRnu})) designed to reproduce the data points, above which the neutrino cooling is efficient, i.e., it is the NDAF area. Below this boundary, the neutrino cooling is insignificant, and it is the ADAF area.}
\label{fig:MdotRnu}
\end{figure}

In a hybrid regime, the inner area of the disk is a NDAF, and the outer area is an ADAF. The boundary in the ($\dot{M}, R$) parameter space between two areas was studied by Chen \& Beloborodov (2007) for the collapsar scenario. From their Figure 11, we find that the following empirical relation can give the NDAF/ADAF transition radius $R_{\nu}$ reasonably well (see Figure \ref{fig:MdotRnu}):
\begin{equation} \label{eq:MdotRnu}
\dot{M}(R_{\nu})= 0.015 \left(\frac{R_{\rm in}}{R_s}\right)^{1.2} \left(1+ \frac{R_{\nu}}{10R_{\rm in}}\right)^2~ M_{\odot} s^{-1},
\end{equation}
where $R_{\rm in}$ is the innermost radius of the disk, and it depends on the BH spin $a_h$: $R_{\rm in}\simeq 3 R_s$ for $a_h= 0$, and $R_{\rm in} \simeq  R_s$ for $a_h= 0.95$. Here, $R_s$ is the Schwarzschild radius of the black hole.

One needs to track the evolution of $R_{\nu}$ in order to know the mass outflow rate $\dot{M}_w(t)$. However, Chen \& Beloborodov's calculation (thus, Equation (\ref{eq:MdotRnu})) was for a constant BH mass of $3 M_{\odot}$, whereas in our calculation of fallback accretion, the BH grows in mass, from $3 M_{\odot}$ to typically $10 M_{\odot}$. We are unaware of any detailed calculation of $R_{\nu}$ for larger BH masses. Noting that all radii in Equation (\ref{eq:MdotRnu}) essentially scale with $M_{BH}$, and generally the boundary accretion rate $\dot{M}(R_{\nu})$ shall increase with $M_{BH}$ (for fixed $R_{\nu}$/$R_{\rm in}$), we propose to generalize Equation (\ref{eq:MdotRnu}) to larger BH masses by
\begin{equation}      \label{eq:MdotRnuG}
\frac{\dot{M}(R_{\nu})}{0.015 M_{\odot} s^{-1}} (\frac{3 M_{\odot}}{M_{BH}}) = \left(\frac{R_{\rm in}}{R_s}\right)^{1.2} \left(1+ \frac{R_{\nu}}{10R_{\rm in}}\right)^2.
\end{equation}
Here, a linearly proportional dependence on $M_{BH}$ is assumed for $\dot{M}(R_{\nu})$, though this needs to be checked through detailed NDAF disk calculation for high $M_{BH}$. Chen \& Beloborodov's calculation shows that $R_{\nu}$'s dependence on the viscosity parameter $\alpha$ can be accurately accounted for by multiplying the right hand side of Equation (\ref{eq:MdotRnuG}) by $(\alpha/0.1)^{5/3}$ (Equation (42) of Chen \& Beloborodov (2007)). As was shown in Figures 11 and 12 of Chen \& Beloborodov (2007), a smaller $\alpha$ will cause the boundary to move down to lower $\dot{M}$ region. Similarly, Kohri et al. (2005) have pointed out that the range of $\dot{M}$ over which the disk wind is inefficient increases for smaller $\alpha$. Following Kohri et al. (2005), in all the following calculation we take $\alpha=0.1$ which is a reasonable value for the collapsar scenario since it has very dynamical environment, strong seed magnetic field and fast MRI growth.

\begin{figure}
\centering
\includegraphics[width=8cm]{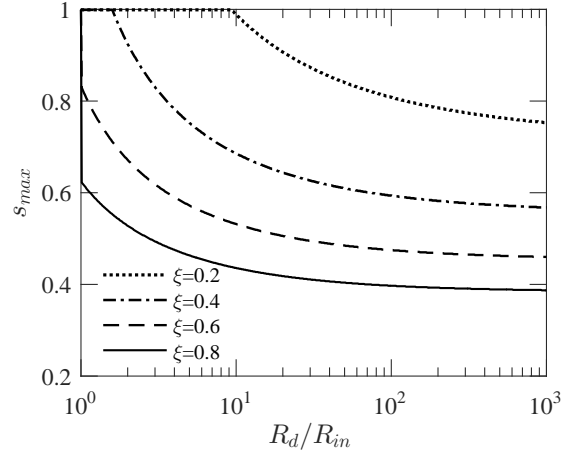}
\caption{ The relationship between the upper limit of $s$ and $R_d/R_{\rm in}$ in the pure ADAF case. With the expansion of the accretion disk, $s$ will decrease.}
\label{fig:Smax}
\end{figure}

Now $R_{\nu}$ can be solved for from Equation (\ref{eq:MdotRnuG}) at any stage of the fallback accretion.
Let $R_d$ be the radial size of the accretion disk. If $R_d \leq R_{\nu}$, the disk is a pure NDAF. If $R_{\nu} \leq R_{\rm in}$, it is a pure ADAF. For the hybrid case (i.e., intermediate accretion rates), $R_{\rm in} < R_{\nu} < R_d$.

Next, we consider the wind energy, for which we adopt the analytically descriptive formulas in
Kohri et al. (2005; their section 3) in the following. In doing this, we take into consideration the
NDAF/ADAF boundary condition (Eq. 2). A crude treatment of the ADAF with wind mass loss is that the accretion rate at radius $R$ is (Kohri et al. 2005)
\begin{equation}\label{accretionrate}
\Mdot(R) = \Mdot_{d} \left(\frac{R}{R_{d}}\right)^{s},
\end{equation}
where $\Mdot_{d}$ is the accretion rate at $R_d$, and the constant index $s$ is between 0 and 1. For the NDAF area in the hybrid regime, one needs to replace $\Mdot_d$ with $\Mdot_{\nu}$ and replace $R_d$ with $R_{\nu}$.

The larger the $s$, the more vigorous the outflow. In the most extreme case, the local accretion rate decreases linearly with $R$ decreasing (going inward) and the disk is in the form of convection-dominated accretion flow (CDAF; Narayan et al. 2000; Quataert \& Gruzinov 2000; Narayan et al. 2001). Pen et al. (2003) and Igumenshchev et al. (2000, 2003) estimated $0.8 \lesssim s \lesssim 1$ by simulations. Yuan et al. (2003) considered $s \sim 0.3$ in their outflow model about Sagittarius A* though that model is in the sub-Eddington ADAF regime. We will show from the energy conservation perspective that, $s$ should have an upper limit below 1.

According to Kohri et al. (2005), one can estimate that $s \approx s_0 \times f_{a}$. Here $s_0$ is a global constant and $f_{a}$ measures the degree of advection of the accretion flow and is approximately the disk height-to-radius ratio $H/R$. Kohri et al. (2005) further show that $f_a \approx 0.5$ at the NDAF/ADAF boundary. In the pure NDAF regime, $f_a \ll 0.1$ (see their Figure 3). In the hybrid regime, there are always
$0.1 \lesssim f_{a} \lesssim 0.5$ for the NDAF area and $0.5 \lesssim f_{a} \lesssim 1$ for the ADAF area.

Following this, we will assume that there is no outflow in the pure NDAF regime (because of negligible $f_a$), but the outflow is still expected in the NDAF region of the hybrid regime. We take a conservative
estimate that in the hybrid regime, $f_a=0.1$ in the NDAF area, $f_{a}=0.5$ in the ADAF area. For the pure ADAF regime, we still take $f_{a}=0.5$ for the entire disk. This is equivalent to say, if $s$ is the mass loss index in the ADAF, then the index in the NDAF region of the hybrid regime is $s_n=s/5$.

Therefore, in the outer ADAF region of the hybrid regime, the mass outflow rate from the annulus $(R,R+dR)$ is
\begin{equation}\label{eq:ADAFmassoutflowrate}
d\Mdot_{w}  = s\frac{\Mdot_d}{R_d}(\frac{R_d}{R})^{1-s}dR.
\end{equation}
while in the inner NDAF area, the differential mass outflow rate is
\begin{equation}\label{eq:NDAFmassoutflowrate}
d\Mdot_{w}  = \frac{s}{5} \frac{\Mdot (R_{\nu})}{R_{\nu}}(\frac{R_\nu}{R})^{1-s/5}dR.
\end{equation}

This mass carries a kinetic energy of
\begin{equation}\label{Ewdot}
d\Edot_{w}=\frac{1}{2} \xi v_{e}^2 d \Mdot_{w},
\end{equation}
where $v_e=2GM(R)/R=c^2 R_s/R$ is the local escape speed squared, and $\xi$ is a numerical factor which Kohri et al. (2005) took to be 0.1 to 1. We will take a conservative value of $\xi=0.3$. Integrating Equation (\ref{eq:ADAFmassoutflowrate}) from $R_{\nu}$ to $R_d$ gives the kinetic power carried by the disk wind from the ADAF region of the hybrid regime
\begin{equation}\label{eq:EwADAFH}
\Edot_{w}=\frac{s R_s \xi \Mdot_{d} c^2}{2(1-s)R_d} [(\frac{R_{d}}{R_\nu})^{1-s} -1].
\end{equation}
Integrating Equation (\ref{eq:NDAFmassoutflowrate}) from $R_{\rm in}$ to $R_{\nu}$ gives the kinetic power carried by the disk wind in NDAF area in the hybrid regime
\begin{equation}\label{eq:EwNDAFH}
\Edot_{w}=\frac{s R_s \xi \Mdot (R_{\nu}) c^2}{10(1-s/5)R_{\nu}} [(\frac{R_{\nu}}{R_{\rm in}})^{1-s/5} -1].
\end{equation}
For the pure ADAF regime, replacing $R_{\nu}$ with $R_{\rm in}$ in Equation (\ref{eq:EwADAFH}) gives the disk wind power in that regime.

From the point of view of energy conservation, the value of $s$ should have an upper limit. The energy carried by the wind is essentially provided by the energy released from the material that eventually goes into the BH. Thus, it shall satisfy: $\Edot_{w} \leq GM \Mdot_h (1/R_{\rm in}-1/R_d)/2$. For the pure ADAF regime, the upper limit of $s$ in this sense depends on $\xi$ and the size of the disk (i.e., $R_d/R_{\rm in}$),  and is plotted in Figure~\ref{fig:Smax}. In our calculation of the disk wind energy, we will take very conservative values $s$=0.15, 0.25, 0.35.

\begin{figure*}
\begin{minipage}[t]{0.5\textwidth}
\centering
\includegraphics[width=8cm]{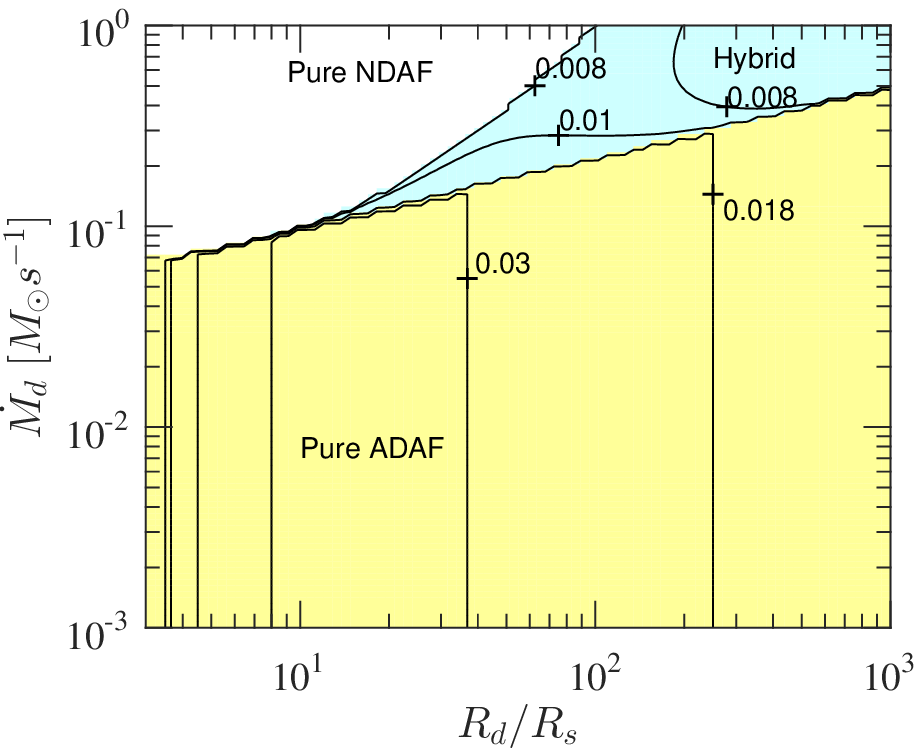}
\end{minipage}
\begin{minipage}[t]{0.5\textwidth}
\centering
\includegraphics[width=8cm]{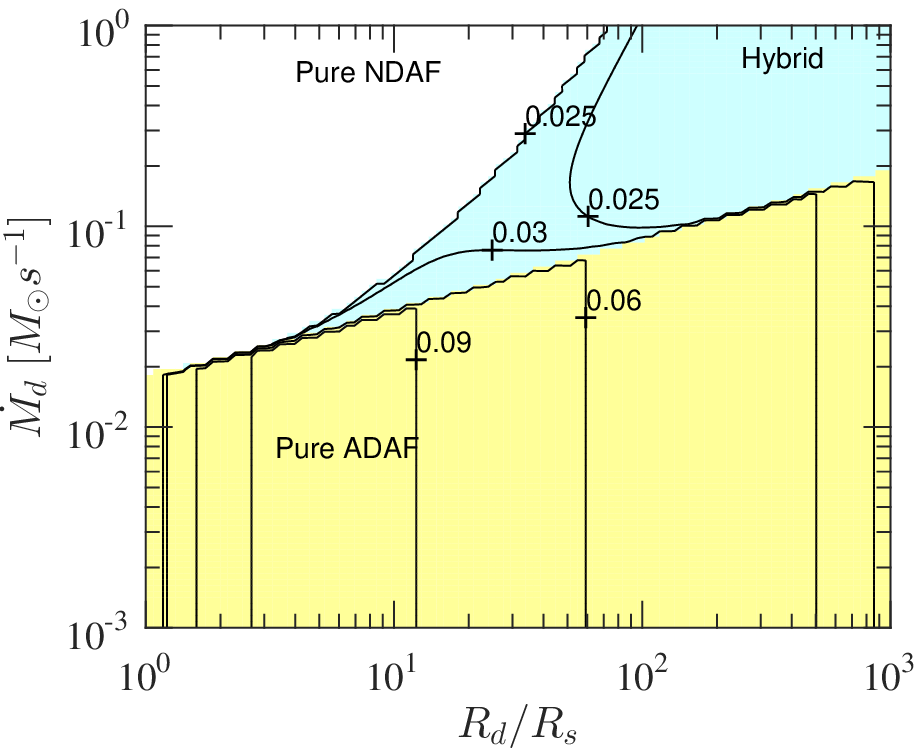}
\end{minipage}
\caption{Contours of the pseudo-efficiency $\dot E_w / (\xi \dot M_d c^2)$ of the disk wind power feedback. Here, $M_{BH}=3M_{\odot}$, $\alpha=0.1$, $s=0.35$, and $a_h=0$ (left), $a_h=0.95$ (right). Yellow areas denote the parameter spaces that the disk is a pure ADAF, and light-blue areas represent that the disks are the mixed states of NDAFs and ADAFs. White areas represent that the disk is a pure NDAF.}
\label{fig:contour}
\end{figure*}

As was pointed out in Kohri et al. (2005), a pseudo-efficiency for the wind power feedback can be represented by $\dot E_w / (\xi \dot M_d c^2)$.
Contours of this quantity, calculated from Equation (\ref{eq:EwADAFH}) and Equation (\ref{eq:EwNDAFH}), are plotted in the Figure~\ref{fig:contour} over the $\dot{M}_d$-$R_d$ plane which is divided into three regimes according to Equation (\ref{eq:MdotRnuG}). It serves to easily show that for any given values of the parameter set ($\xi$, $a_h$, $R_d$, $\Mdot_d$), which accretion regime it is in, and how much kinetic power of the wind is available. From Equation (\ref{eq:EwADAFH}), it is
obvious that for the pure ADAF regime this pseudo-effciency does not depend on $\Mdot_d$. Also, as was shown in Figure 7 of Kohri et al. (2005), with the increase of $R_d/R_s$, it increases first and then decreases.

The instantaneous disk wind energy is approximately given by
\begin{equation}\label{Windenergy}
E_{w} (t) =(t-t_o)\times \dot{E}_{w} (t),
\end{equation}
where $t_o$ is the time of outflow appearance, i.e., when the hybrid regime firstly appears.

\section{The initial explosion, mass fallback and disk formation}
Here we treat the initial explosion and mass fallback in a simplified but physically concise manner. During the collapse of the star, the formation of a NS sends out an initial explosive shockwave of energy $E_{ini}$, which can unbind only the outermost part of the envelope. Here, we assume that all the envelope material obtain a uniform outward velocity $v_{ini}=(2E_{ini}/M_{env})^{1/2}$, where $M_{env}$ is all the mass exterior to a 2.1 $M_{\odot}$ proto-NS. Then $R_b=2GM(R_b)/{v_{ini}}^2$ would be the \emph{initial} radius in the pre-collapse star that separates the inner bound region and the outer unbound region.

Consider the material that was initially at $R_{ini} < R_b$ before the collapse. During the initial explosion, it will move outward firstly to $R_{fb}$, then fall back, where $R_{fb}$ is simply related to the initial radius $R_{ini}$  in pre-SN models as
\begin{equation}\label{R_n}
\frac{1}{R_{fb}}=\frac{1}{R_{ini}}-\frac{{v_{ini}}^2}{2GM(R_{ini})}.
\end{equation}

With fallback accretion, the proto-NS collapses to a BH whose initial mass we assume to be $ 3 M_\odot$. As the subsequent matter falls into the BH, the total angular momentum $J$ of the BH increases, which changes its Kerr parameter $a_h=J/M$ (here $ \rm G= \rm c=1$). For the falling matter to form a disk around the BH, its specific angular momentum must exceed the specific angular momentum at the innermost stable circular orbit (ISCO)
\begin{equation}
j_{isco} = \frac{(GMR_{\rm in})^{1/2} \left[R_{\rm in}^2-a_hR_s(R_{\rm in}R_s/2)^{1/2}+a_h^2R_s^2/4 \right]} {R_{\rm in}\left[R_{\rm in}^2-3R_{\rm in}R_s/2+a_hR_s(R_{\rm in}R_s/2)^{1/2}
\right]^{1/2}}\;,
\label{eq:jGR}
\end{equation}
otherwise the material falls into the BH directly. By angular momentum conservation, the size of the disk is $R_{d} = {j}^2(R_{ini})/ {GM(R_{ini})}$, where $j(R_{ini})$ is the specific angular momentum distribution in pre-SN models. We estimate the mass fallback rate by
\begin{equation}\label{mdot}
\Mdot_{fb} = \frac{M_{i+1}-M_i}{t_{i+1}-t_i},
\end{equation}
where $M_i$ is the enclosed mass at a certain radius in pre-SN models, and $t_i \backsimeq [R_{fb}^3 / GM(R_{ini})]^{1/2}$ is the free fall time from $R_{fb}$ to $R_d$.
Since in the collapsar scenario, the disk accretion time scale is much shorter than $t_i$, the disk accretion rate $\Mdot(R_d)$ can be safely take to be equal to $\Mdot_{fb}$.

Next, we estimate the gravitational binding energy of the infalling envelope $E_b$, which we will compare with the wind energy $E_w$ (Equation (\ref{Windenergy})). After the initial expansion of the envelope, the material has expanded to its maximum radius $R_{fb}$, and is ready to fall back. At a given time $t$ for an mass element $dM$ whose initial radius is $R_{ini}$,
\begin{equation}\label{bindingenergy}
d E_{b}(t) = \frac{GM(R_{ini}) dM(R_{ini})}{R_{fb}-\frac{GM(R_{ini})t^2}{{R_{fb}}^2}+R_d}.
\end{equation}
Here $R_d$ is added to the denominator in order to avoid it from being zero. And then we have the binding energy:
\begin{equation}\label{intEwdot}
E_b (t)= \int_{M(R_{ini})}^{M(R_b)} \frac{dE_b(t)}{dM(R_{ini})} dM(R_{ini}).
\end{equation}

\section{Applying to pre-supernova models}

We would like to use a wide range of pre-SN models to be the initial setup of our disk wind model, so that the model can cover as many situations as it could. We use the pre-SN models from Perna et al. (2014) computed by MESA code. These pre-SN models cover a wide range of their ZAMS parameter space: initial mass  $\sim 13-40M_\odot$, initial surface rotational velocities $\sim 25\%-75\% $ of the critical velocity (e.g., Equation (1) of Perna et al. 2014), and metallicities of 1\%,10\% and 100\% of the solar value. We do not consider the influence of magnetic field for now. The mass of the core below $\sim 2.1 M_\odot$ will become a neutron star first, but most of these pre-SN models will form a BH with fallback accretion for a large range of explosion energies, as was shown by Figure 6 of Perna et al. (2014).

\begin{figure*}
\begin{minipage}[t]{0.5\textwidth}
\centering
\includegraphics[width=8cm]{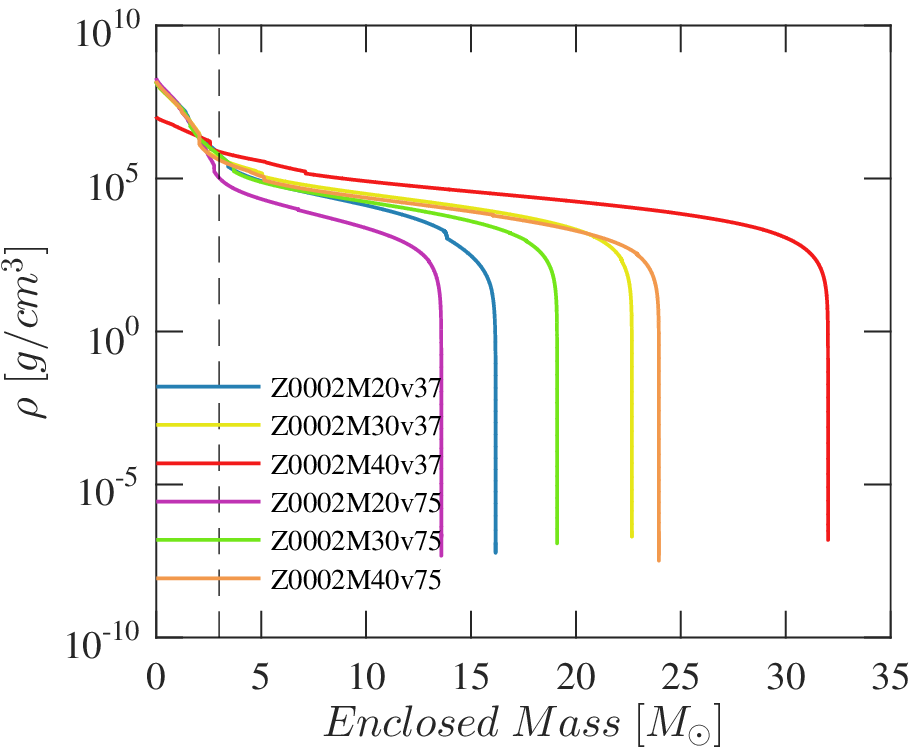}
\end{minipage}
\begin{minipage}[t]{0.5\textwidth}
\centering
\includegraphics[width=8cm]{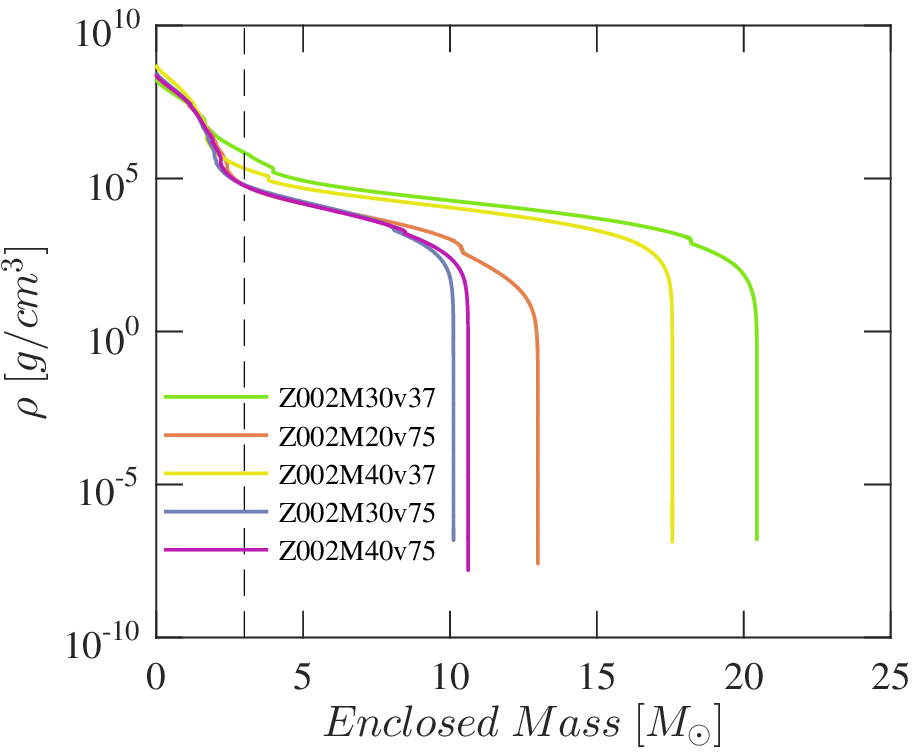}
\end{minipage}
\begin{minipage}[t]{0.5\textwidth}
\centering
\includegraphics[width=8cm]{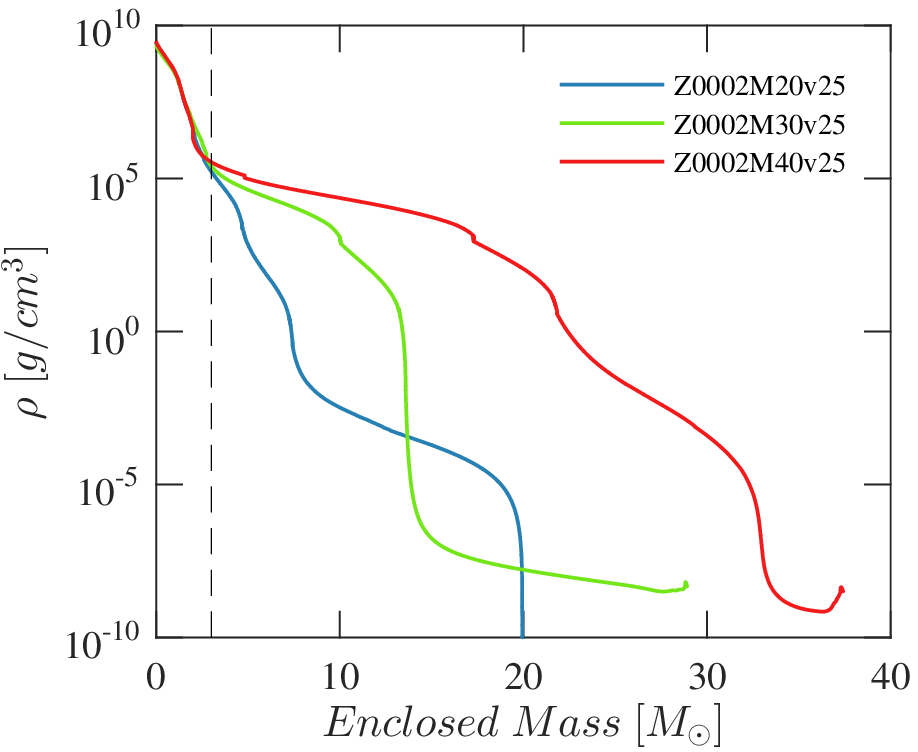}
\end{minipage}
\begin{minipage}[t]{0.5\textwidth}
\centering
\includegraphics[width=8cm]{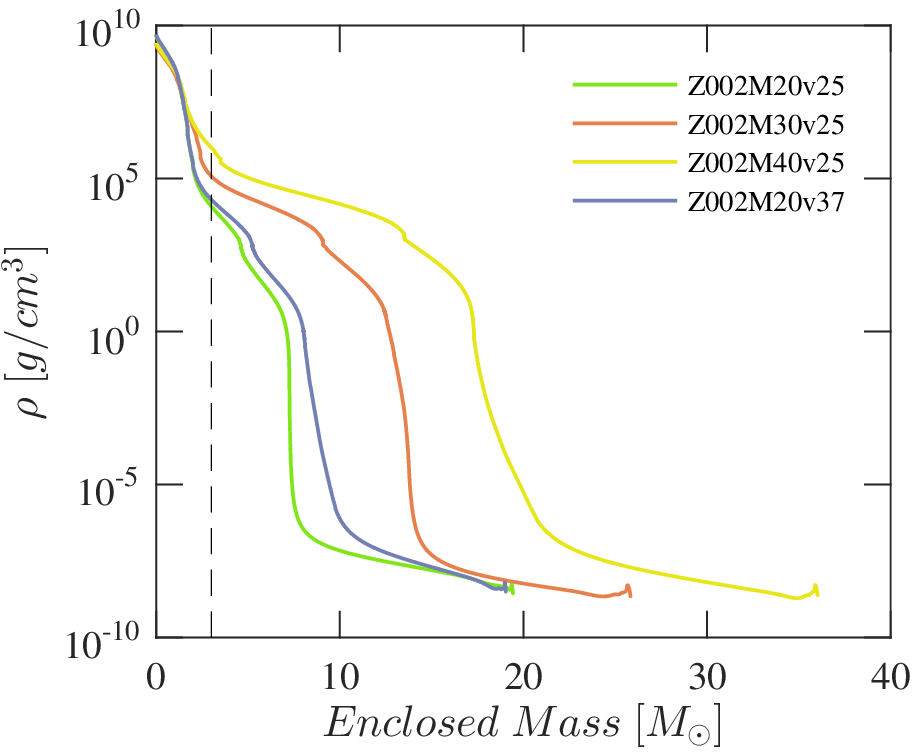}
\end{minipage}
\linespread{1}
\caption{The density profiles of pre-SN models (Perna et al. 2014) that we used. Vertical straight lines represent the initial mass of the BH we set. Top left (right) panel: rapid accretion cases with metallicity about 1\%(10\%) of the solar value. Lower left (right) panel: long-lasting accretion cases with metallicity about 1\% (10\%) of the solar value.
The naming of pre-SN models is such that the zero-age main sequence star of Z0002M20v75 has a surface velocity is $75\%$ of critical value and an initial mass is $20M_\odot$, and its metallicity is 1\% of the solar value.}
\label{fig:Densityprofile}
\end{figure*}

\begin{figure*}
\begin{minipage}[t]{0.5\textwidth}
\centering
\includegraphics[width=8cm]{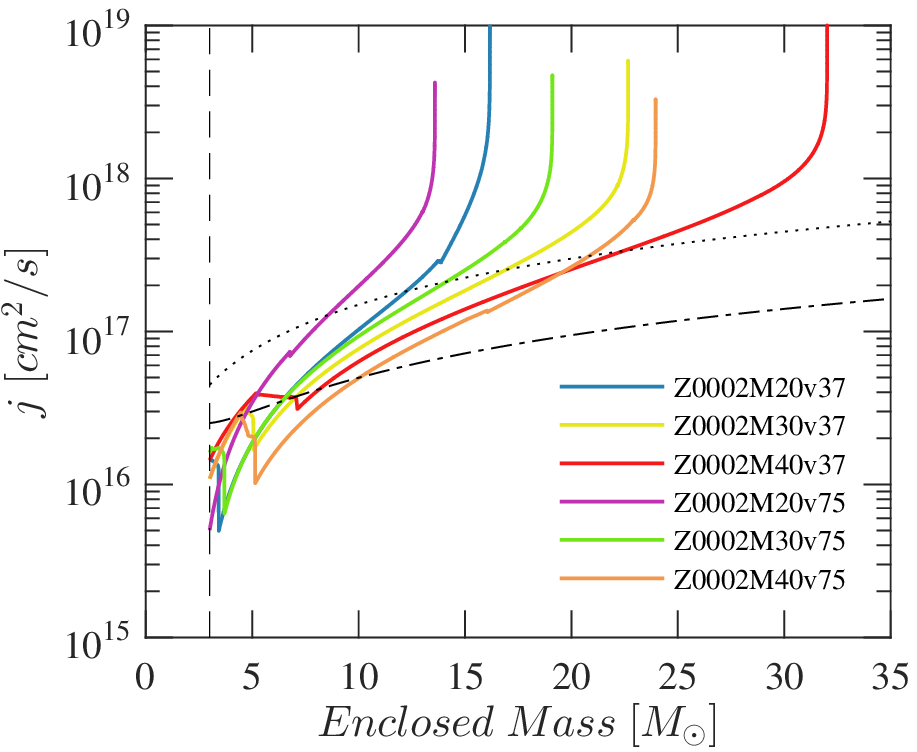}
\end{minipage}
\begin{minipage}[t]{0.5\textwidth}
\centering
\includegraphics[width=8cm]{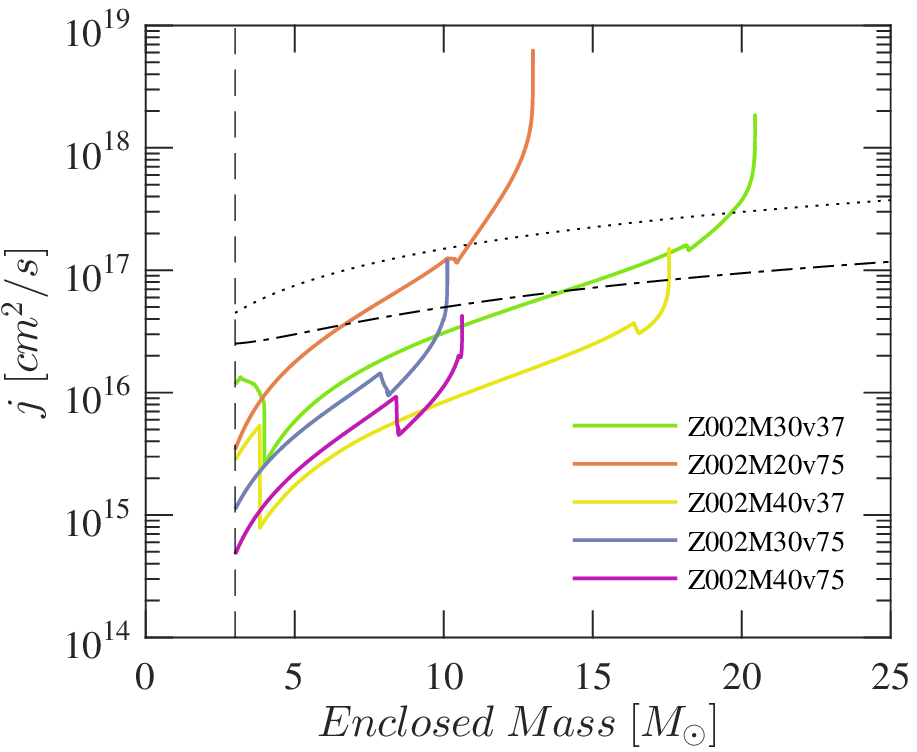}
\end{minipage}
\begin{minipage}[t]{0.5\textwidth}
\centering
\includegraphics[width=8cm]{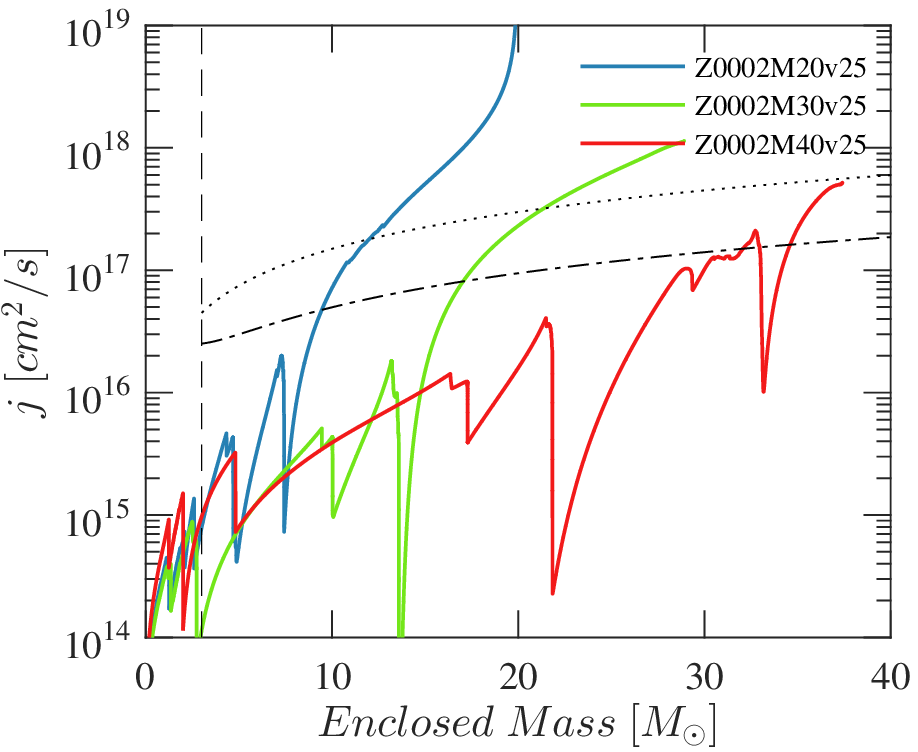}
\end{minipage}
\begin{minipage}[t]{0.5\textwidth}
\centering
\includegraphics[width=8cm]{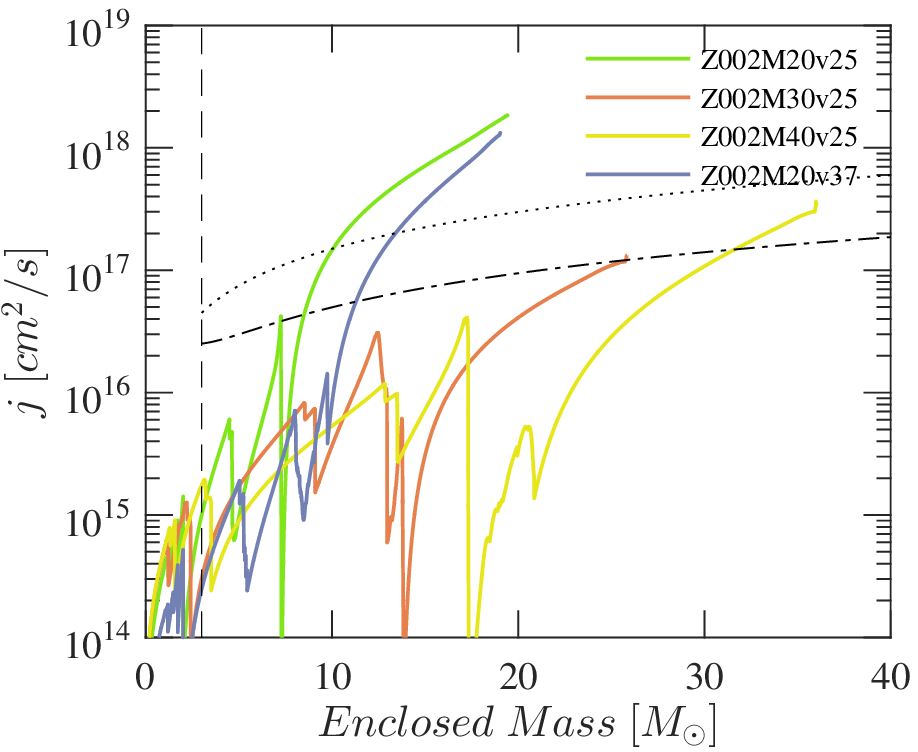}
\end{minipage}
\linespread{1}
\caption{Specific angular momentum distribution in pre-SN models. Vertical straight lines represent the initial mass of the BH we set. The dotted (dotted-dashed) line represents $j_{isco}$ with BH masses equal to enclosed mass and $a_h=0.1$ (0.95). The arrangement of panels is the same as in Figure~\ref{fig:Densityprofile}.}
\label{fig:AM}
\end{figure*}

We choose a sub-grid of these pre-SN models, with initial masses of $20M_\odot$, $30M_\odot$, $40M_\odot$, surface velocities of $ 25\%$, $ 37\%$, $75\% $ of the critical velocity, and metallicities of 1\%,10\% of the solar value. Figure~\ref{fig:Densityprofile} and Figure~\ref{fig:AM} plot the density profiles and the distributions of specific angular momentum of the each pre-SN models, respectively.

According to the outcome of our fallback calculation, these pre-SN stars can be divided into two categories. The first is rapid fallback: the free fall times of the outermost material are $10^2$ - $10^3$s. The other case is long-lasting fallback: the free-fall times of the outermost material are on the order of $10^7s$.
For the former category, if the initial explosion energy is not too large($\le$ a few $\times 10^{51}$erg), almost all the envelope will fall back. For the latter category, because of their large radius, the binding energy of the outermost layer is very low, which means a weak explosion will unbind the outermost layer. This will happen if the progenitor is a RSG (Nadezhin 1980; Lovegrove \& Woosley 2013; Kashiyama \& Quataert 2015). The third column of Tables 1 - 3 shows the mass of unbound matter in our calculations for 3 values of initial explosion energy.

We note that if the initial explosion is anisotropic, a considerable of energy will be deposited along the rotation axis, with a relatively low amount of energy release to the equatorial area, which makes most of the envelope remains bound and falls back even if the initial explosion is energetic (Hillebrandt \& H\"oflich 1989; Spyromilio 1991; McCray 1993; Woosley et al. 1994; Fassia et al. 1998; Perna et al. 2014).

\begin{figure*}
\begin{minipage}[t]{0.5\textwidth}
\centering
\includegraphics[width=8cm]{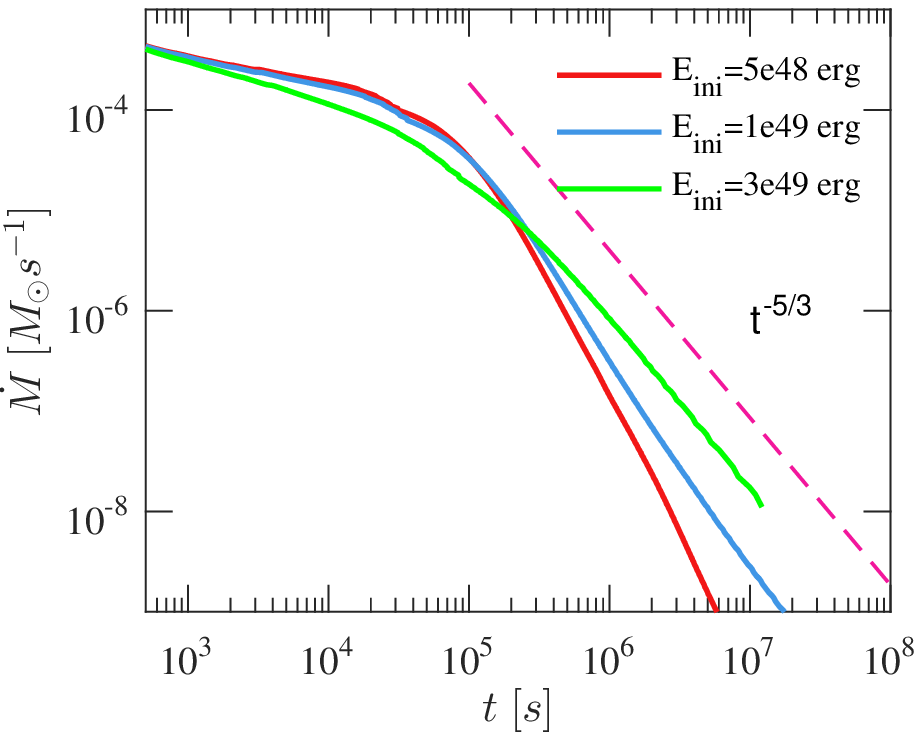}
\end{minipage}
\begin{minipage}[t]{0.5\textwidth}
\centering
\includegraphics[width=8cm]{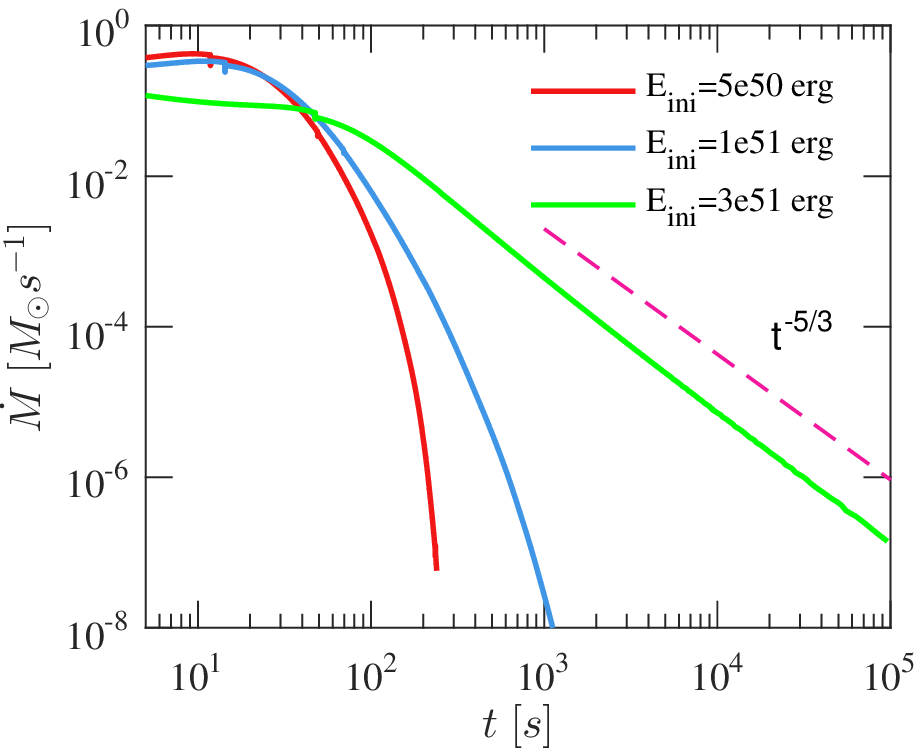}
\end{minipage}
\caption{Fallback rates of pre-SN models Z0002M20v25 (left) and Z0002M20v75 (right). Even for a relatively low initial explosion energy ($E_{ini}=5 \times 10^{49}$erg), there is about $6.3M_{\odot}$ of the envelope unbound for Z0002M20v25. Hence, for long-lasting accretion cases, a weak explosion will unbind the outer envelope.}
\label{fig:fallbackrates}
\end{figure*}

Figure~\ref{fig:fallbackrates} shows the fallback rate history with different explosion energies $E_{ini}$ for pre-SN models Z0002M20v25 (a long-lasting one) and Z0002M20v75 (a rapid fallback example). If the initial explosion is energetic enough, the outer layer of the envelope will be unbound and the tail of the fallback rate $\dot M_{fb}$ reproduces the canonical power-law $~t^{-5/3}$ (Michel 1988; Chevalier 1989; Dexter \& Kasen 2013; Zhang et al.2008).

Heger et al. (2003) show that for ZAMS masses larger than about $40M_{\odot}$ with metallicities lower than the solar value, BH can form directly in the collapse. Therefore, for rapid fallback cases with $M=40M_\odot$, we assume that the core can collapse to BH directly.
For all other cases, we take three initial explosion energies $E_{ini}=10^{49}$erg, $10^{50}$erg and $10^{51}$erg in our calculation.

\section{results}

\linespread{0.8}
\begin{table*}\centering
\setlength{\tabcolsep}{5mm}{
\begin{tabular}{cccccc}
\hline
\hline
& Pre-SN stellar& Unbound mass after    &  $M_{BH}(t_w)$  &  $M_{enve}(t_w)$ & $E_w(t_w)$ \\
&mass [$\Msun$] & NS formation [$\Msun$]  & [$\Msun$]    & [$\Msun$]  & [$10^{51}$erg]  \\
\hline
\hline
  & & \; \; \; Rapid  &           &     \\
\hline
Z0002M20v37 &16.2& 0  &     13.0\;-\;13.3   &       2.75\;-\;3.10   &    4.62\;-\;5.62\\
\hline
Z0002M30v37 &22.7& 0  &     18.8\;-\;19.5   &       3.08\;-\;3.68   &    8.89\;-\;10.8\\
\hline
Z0002M40v37 &32.0& -  &     26.4\;-\;27.5   &       4.33\;-\;5.26   &    14.6\;-\;18.4\\
\hline
Z0002M20v75 &13.6& 0  &     9.56\;-\;9.96   &       3.57\;-\;3.92   &    4.91\;-\;5.40\\
\hline
Z0002M30v75 &19.1& 0  &     15.5\;-\;16.1   &       2.93\;-\;3.42   &    7.10\;-\;8.49\\
\hline
Z0002M40v75 &24.0& -  &     20.4\;-\;21.0   &       2.87\;-\;3.46   &    8.29\;-\;10.2\\
\hline
Z002M30v37  &20.5& 0  &     19.4\;-\;19.4   &       0.99\;-\;1.06   &    2.14\;-\;2.38\\
\hline
Z002M20v75  &13.0& 0  &     10.0\;-\;10.1   &       2.85\;-\;2.93   &    2.61\;-\;2.85\\
\hline
Z002M40v37  &17.6& -  &            17.6          &            ND          &         ND       \\
\hline
Z002M30v75  &10.1&0   &            10.1          &            ND          &         ND       \\
\hline
Z002M40v75  &10.6& -  &            10.6          &            ND          &         ND       \\
\hline
\hline
  &  &\; \; \; Long-lasting &           &     \\
\hline
Z0002M20v25 &19.9&0.04 &   11.6\;-\;11.6   &       8.33      &    0.30\;-\;0.58\\
\hline
Z0002M30v25 &28.9&13.9 &          15.0          &          ND        &         ND       \\
\hline
Z0002M40v25 &37.4&3.87  &          33.5          &          ND        &         ND       \\
\hline
Z002M20v25  &19.4&11.8 &          7.61          &          ND        &         ND       \\
\hline
Z002M30v25  &25.8&11.4 &          14.4          &          ND        &         ND       \\
\hline
Z002M40v25  &36.0&12.9 &          23.1          &          ND        &         ND       \\
\hline
Z002M20v37  &19.0&9.17  &          9.87          &          ND        &         ND       \\
\hline
\end{tabular}}
\linespread{1}
\caption{\upshape The results of our calculations of each pre-SN models with initial explosion energy $E_{ini}=10^{49}$erg.
(1) The first column is the mass of the star at the pre-SN stage.
(2) The second column indicates the mass of unbound matter during the initial explosion following the NS formation. It does not apply to rapid accretion models with initial mass $M=40M_\odot$ because the cores will collapse to BHs directly. For long-lasting fallback cases, we also consider the initial explosion even if $M=40M_\odot$.
(3) The third column is the mass of the BH at $t_w$. If there is always no disk formation (abbreviated as ND in the fourth and fifth columns), then this data is the final remnant BH mass.
(4) The fourth column is the mass of the material be ejected by disk wind.
(5) The fifth column is the energy of the disk wind at the explosion point. The numerical interval in each data corresponds to $s=0.15-0.35$.}
\end{table*}

\linespread{0.8}
\begin{table*}\centering
\setlength{\tabcolsep}{5mm}{
\begin{tabular}{cccccc}
\hline
\hline
& Pre-SN stellar& Unbound mass after    &  $M_{BH}(t_w)$  &  $M_{enve}(t_w)$ & $E_w(t_w)$ \\
&mass [$\Msun$] & NS formation [$\Msun$]  &  [$\Msun$]    & [$\Msun$]  & [$10^{51}$erg]  \\
\hline
\hline
  & & \; \; \; Rapid   &          &     \\
\hline
Z0002M20v37 &16.2& 0    &   12.9\;-\;13.3   &       2.79\;-\;3.14   &    4.65\;-\;5.64\\
\hline
Z0002M30v37 &22.7& 0    &   18.7\;-\;19.4   &       3.12\;-\;3.72   &    8.84\;-\;10.7\\
\hline
Z0002M40v37 &32.0& -    &   26.4\;-\;27.5   &       4.33\;-\;5.26   &    14.6\;-\;18.4\\
\hline
Z0002M20v75 &13.6& 0    &   9.50\;-\;9.90     &       3.63\;-\;3.98   &    4.86\;-\;5.31\\
\hline
Z0002M30v75 &19.1& 0    &   15.5\;-\;16.1   &       2.96\;-\;3.46   &    7.14\;-\;8.39\\
\hline
Z0002M40v75 &24.0& -    &   20.4\;-\;21.0   &       2.87\;-\;3.46   &    8.29\;-\;10.2\\
\hline
Z002M30v37  &20.5& 0    &   19.4\;-\;19.4   &       0.99\;-\;1.06   &    2.11\;-\;2.29\\
\hline
Z002M20v75  &13.0& 0    &   10.0\;-\;10.1   &       2.86\;-\;2.93   &    2.56\;-\;2.85\\
\hline
Z002M40v37  &17.6& -     &         17.6          &            ND          &         ND       \\
\hline
Z002M30v75  &10.1&0      &         10.1          &            ND          &         ND       \\
\hline
Z002M40v75  &10.6& -     &         10.6          &            ND          &         ND       \\
\hline
\hline
  & &\; \; \; Long-lasting    &        &     \\
\hline
Z0002M20v25 &19.9&10.3 &          9.64          &          ND        &         ND       \\
\hline
Z0002M30v25 &28.9&15.1 &          13.8          &          ND        &         ND       \\
\hline
Z0002M40v25 &37.4&5.01  &          32.4          &          ND        &         ND       \\
\hline
Z002M20v25  &19.4&12.1 &         7.31           &          ND        &         ND       \\
\hline
Z002M30v25  &25.8&12.1 &          13.8          &          ND        &         ND       \\
\hline
Z002M40v25  &36.0&16.6 &          19.4          &          ND        &         ND       \\
\hline
Z002M20v37  &19.0&10.3 &          8.75          &          ND        &         ND       \\
\hline
\end{tabular}}
\linespread{1}
\caption{\upshape The same as in Table 1, but the initial explosion energy is $E_{ini}=10^{50}$erg.
}
\end{table*}

\linespread{0.8}
\begin{table*}\centering
\setlength{\tabcolsep}{5mm}{
\begin{tabular}{cccccc}
\hline
\hline
& Pre-SN stellar& Unbound mass after    &  $M_{BH}(t_w)$  &  $M_{enve}(t_w)$ & $E_w(t_w)$ \\
&mass [$\Msun$] & NS formation [$\Msun$]  &  [$\Msun$]    & [$\Msun$]  & [$10^{51}$erg]  \\
\hline
\hline
  & & \; \; \; Rapid   &          &     \\
\hline
Z0002M20v37 &16.2& 0    &   12.6\;-\;13.0   &       3.13\;-\;3.49   &    4.65\;-\;5.51\\
\hline
Z0002M30v37 &22.7& 0    &   18.5\;-\;19.2   &       3.38\;-\;3.98   &    8.87\;-\;10.7\\
\hline
Z0002M40v37 &32.0& -    &   26.4\;-\;27.5   &       4.33\;-\;5.26   &    14.6\;-\;18.4\\
\hline
Z0002M20v75 &13.6& 0    &   8.79\;-\;9.14     &       4.40\;-\;4.72   &    4.10\;-\;4.36\\
\hline
Z0002M30v75 &19.1& 0    &   15.1\;-\;15.7   &       3.32\;-\;3.81   &    7.06\;-\;8.27\\
\hline
Z0002M40v75 &24.0& -    &   20.4\;-\;21.0   &       2.87\;-\;3.46   &    8.29\;-\;10.2\\
\hline
Z002M30v37  &20.5& 0    &   19.3\;-\;19.4   &       1.02\;-\;1.08   &    1.74\;-\;1.94\\
\hline
Z002M20v75  &13.0&0.07  &   10.0\;-\;10.1   &       2.84\;-\;2.90   &    1.65\;-\;1.71\\
\hline
Z002M40v37  &17.6& -     &         17.6          &            ND          &         ND       \\
\hline
Z002M30v75  &10.1&2e-5  &         10.1          &            ND          &         ND       \\
\hline
Z002M40v75  &10.6& -     &         10.6          &            ND          &         ND       \\
\hline
\hline
  & &\; \; \; Long-lasting    &        &     \\
\hline
Z0002M20v25 &19.9&12.8 &          7.17          &          ND        &         ND       \\
\hline
Z0002M30v25 &28.9&15.3 &          13.6          &          ND        &         ND       \\
\hline
Z0002M40v25 &37.4&12.4 &          25.0          &          ND        &         ND       \\
\hline
Z002M20v25  &19.4&12.9 &          6.55          &          ND        &         ND       \\
\hline
Z002M30v25  &25.8&12.8 &          13.0          &          ND        &         ND       \\
\hline
Z002M40v25  &36.0&18.2 &          17.8          &          ND        &         ND       \\
\hline
Z002M20v37  &19.0&11.4 &          7.61          &          ND        &         ND       \\
\hline
\end{tabular}}
\linespread{1}
\caption{\upshape The same as in Table 1, but the initial explosion energy is $E_{ini}=10^{51}$erg.
}
\end{table*}

Tables 1, 2 and 3 list our results of the calculations of all the pre-SN models with different initial explosion energies.
One may naively expect that a disk should form if the explosion is very weak. However, as we discussed in Section 4, a low $E_{ini}$ will unbind the outer layers for RSG progenitors which correspond to the long-lasting fallback category in Tables 1 - 3. For them, even for $E_{ini}=10^{49}$erg, all the high $j$ layers are unbound. That is why there is no disk formation for most of the long-lasting cases.
Although we have calculated all the pre-SN models, due to the length of the paper, in Figure 8 we plot only 4 of these 9 pre-SN models which can form a disk: Z0002M20v25, Z0002M20v75, Z0002M40v37 and Z002M30v37.

\begin{figure*}
\begin{minipage}[t]{0.5\textwidth}
\centering
\includegraphics[width=8cm]{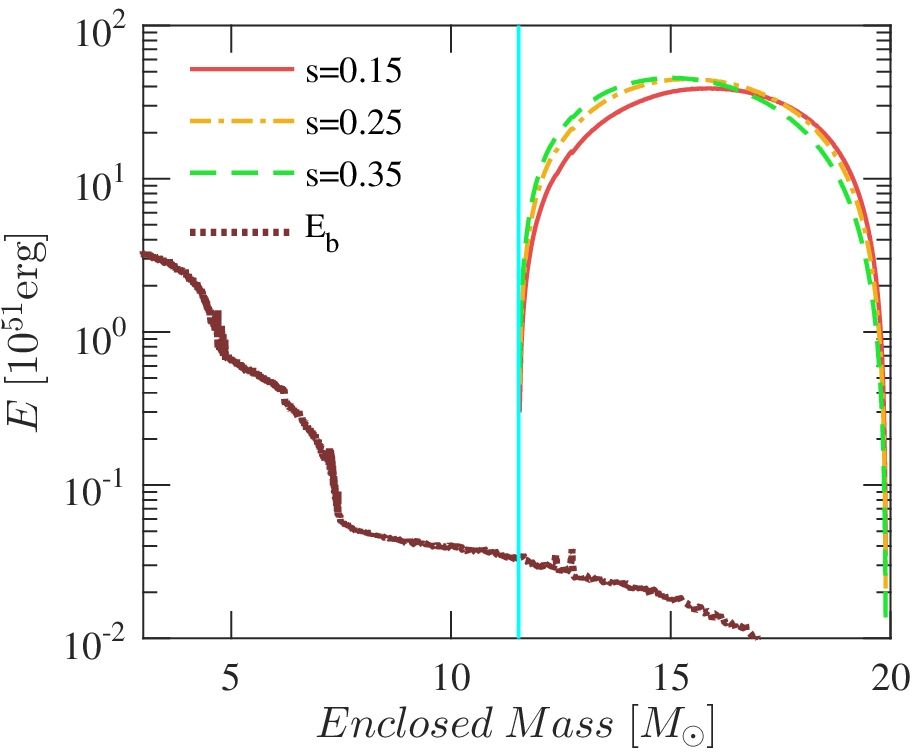}
\end{minipage}
\begin{minipage}[t]{0.5\textwidth}
\centering
\includegraphics[width=8cm]{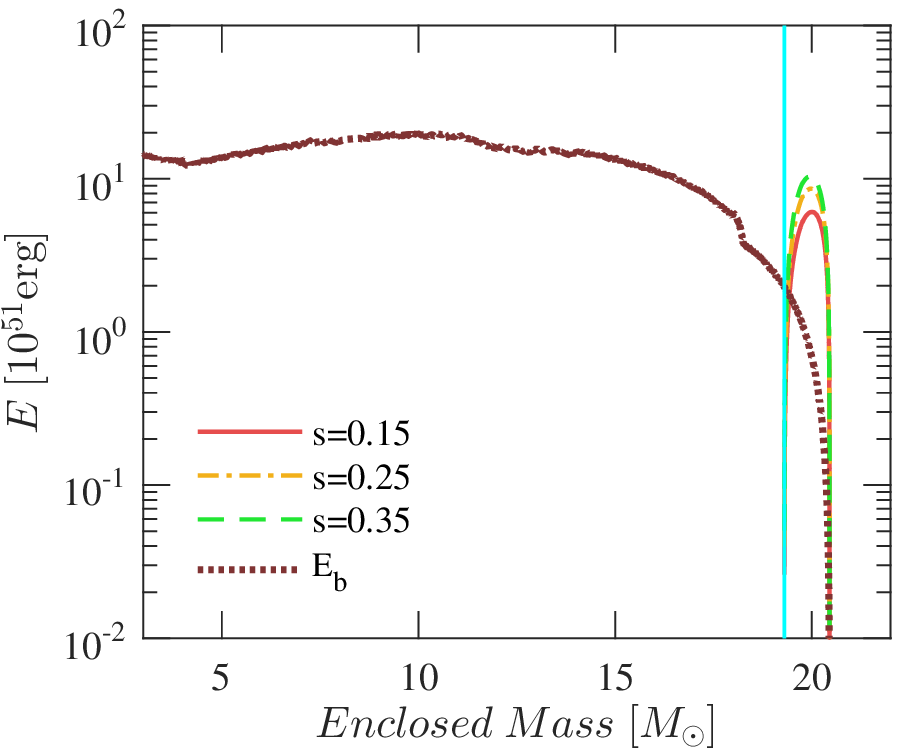}
\end{minipage}
\begin{minipage}[t]{0.5\textwidth}
\centering
\includegraphics[width=8cm]{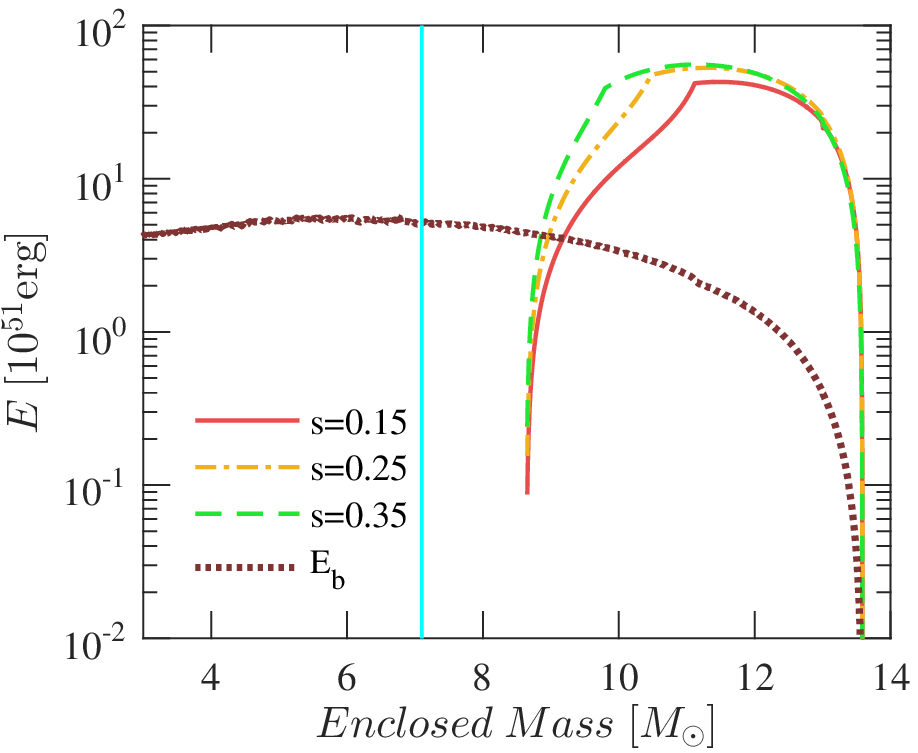}
\end{minipage}
\begin{minipage}[t]{0.5\textwidth}
\centering
\includegraphics[width=8cm]{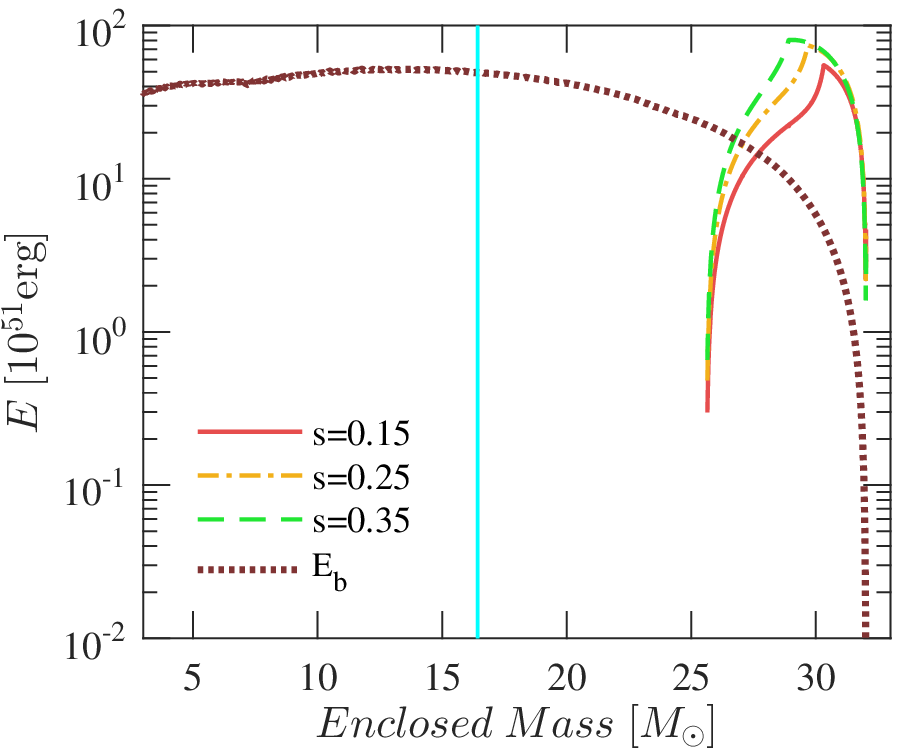}
\end{minipage}
\linespread{1}
\caption{The energies of our calculation using pre-SN models Z0002M20v25 with $E_{ini}=10^{49}$ erg (top left panel) and Z002M30v37 with $E_{ini}=10^{51}$ erg (top right panel), Z0002M20v75 with $E_{ini}=10^{51}$ erg (lower left panel) and Z0002M40v37 with $E_{ini}=10^{51}$ erg (lower right panel). Vertical straight lines identify the time of accretion disk formation. The dotted line represents the binding energy of the infalling envelope. The solid line, dotted-dashed line, and dashed line represent the disk wind energy with different values of $s$. We adopt $\xi=0.3$ in all. Define $t_w$ as the time when $E_w$ exceeds $E_b$ for the first time. Around $t_w$ the disk wind is energetic enough to reverse the collapse of the envelope and unbind it. Thus, we do not need to care about $E_w$ and $E_b$ after $t_w$. In the bottom panels, there is a delay of the wind launching after the disk formation because of the existence of the pure NDAF regime.}
\label{fig:preSN2}
\end{figure*}

Define $t_w$ as the time that $E_w$ exceeds $E_b$ for the first time; it would be the time when the disk wind is energetic enough to reverse the collapse of the envelope and unbound it.
The masses of the envelope that will be unbound, $M_{enve}(t_w)$, and the total energy of the disk wind $E_w(t_w)$ of each pre-SN model in the two categories are also shown in Tables 1 - 3. They will become the ejecta mass and the total energy, respectively, for an observed wind-driven SN explosion. It shows that, for the long-lasting fallbacks, a large amount of the material will be unbound after NS formation even if the initial explosions are weak ($E_{ini}=10^{49}$erg).  The traditional explosion mechanism faces no difficulty for these cases. Thus, we will not address these models further.

For rapid fallback cases,  after the disk formation, the accretion flow is initially a NDAF. With such high accretion rate, the black hole grows fast. The accretion rate eventually drops and the energy of the disk wind accumulates quickly. The total energy of the disk wind is always larger than $10^{51}$erg, even up to $2 \times 10^{52}$erg for some cases.
The huge kinetic energy carried by disk wind will generate a shock wave that expands outward and eventually unbinds the whole envelope. We verified this by calculating the outward propagation of this shock following a procedure similar to Matzner \& Mckee (1999).

The accretion history and subsequently the wind energy budget are mainly determined by three factors. The first is the distribution of specific angular momentum. The second is the initial explosion. These two factors both determine whether and when the disk forms.  The third factor is the compactness of the pre-SN star. It determines the existence and duration of pure NDAF regime, Hybrid regime, and pure ADAF regime. Following O'Connor \& Ott (2011) and Ugliano et al. (2012), a compactness parameter can be defined as: $\zeta_6 \equiv 6000km/R(M_6)$, where $R(M_6)$ is the radius that encloses $M=6M_{\odot}$ at the pre-SN stage.

According to our calculations, the larger $\zeta_6$ is, the longer the duration of the pure NDAF regime. There are exceptions (Z0002M40v75, Z002M30v37) because their angular momenta are low so that the disk is formed late. With a small $\zeta_6$ (e.g., long-lasting model Z0002M20v25), the pure NDAF regime will not appear, as well as the hybrid regime. In this situation, the disk is a pure ADAF as soon as it forms.

In addition, a larger $s$ or $E_{ini}$ tends to make the explosion occur earlier and lead to a larger ejecta mass (see Figure 8 or Tables 1 - 3).

\section{Conclusion}
We presented an analytical, time-dependent accretion disk wind model, for the BH fallback accretion during the stellar core collapse. We applied this model to a series of realistic pre-SN stars, with different initial masses, rotation velocities and metallicities, to assess the feasibility of wind-driven SN explosions.
We found that for some progenitors which satisfy certain conditions, they will be exploded successfully by disk wind.

\begin{enumerate}
\item With relatively low metallicity and relatively large specific angular momentum, the rapid fallback pre-SN stars are the most promising progenitors to be exploded successfully by disk wind.
    In the pre-SN models that we studied, those are the ones with initial v $ \gtrsim 37\%$ of the critical velocity, and $Z \lesssim 1\%$ of the solar value.
If the metallicity is high ($ Z \gtrsim 10\%$ of the solar value), most of the high $j$ layers would have been lost in stellar winds, and there may be no disk formation.

\item Notice that the more massive the progenitor is, the larger the disk wind energy will be.  Also, the mass of the ejected matter will tend to be larger. Hence, a more massive progenitor may produce a more energetic explosion.

\item The total energies of the wind-driven explosions are always larger than $10^{51}$erg and up to $10^{52}$erg.
The traditional model had difficulty in such situations:
(1) Not easy to explain such large explosion energy like in hypernovae (e.g., Woosley \& Bloom 2006; Janka 2012; Janka et al. 2016); (2) The more massive and compact the progenitor is, the more difficult for the star to exploded (e.g., Janka et al. 2007).
Here, the disk wind model can solve these problems.

\item If the initial explosion after NS formation is weak enough in long-lasting cases, then most of the material will remain bound and falls back. Since the exact value of the initial explosion energy is still an open question (e.g., Janka 2012; Perna et al. 2014), such a very weak explosion (shock) may exist and cause the most of the envelope to fall back (Woosley \& Heger 2012; Perna et al. 2014). Under this premise, the disk wind may also cause successful explosions for supergiants with low metallicities, relatively large specific angular momentum. One of the pre-SN models that we calculated, Z0002M20v25, is such an example.
\end{enumerate}

\section{Discussion}

When we consider the accretion flow, we assumed that
there is no outflow when the disk is a pure NDAF. This is of course a crude treatment, and it likely underestimates the promptness of the wind-driven explosion because there may also be some outflow for a pure NDAF (Kohri et al.
2005). Thus, the wind-driven explosion energy and ejecta mass might be even higher than what we presented here if the wind in the pure NDAF regime is properly taken into account.

Another uncertainty resides with the $s$ and $\xi$ parameters. We have considered a conservative range of $s \sim 0.15-0.35$, and $\xi=0.3$. Larger values of these would certainly increase $E_w$ (Equation (\ref{eq:EwADAFH})(\ref{eq:EwNDAFH})).

In this work we considered the energy of the wind only. Jets may exist when the disk is a NDAF.
Jets will deposit their energy into the stellar envelope if they can not break out of the progenitor (e.g., Bromberg et al. 2011) and may drive stellar explosions (e.g., Lazzati et al. 2012).
It will increase the total energy output and the ejecta mass if one takes into account the jet's sideways energy deposition during its penetration through the stellar envelope.

Most massive stars live in binary or multiple systems, and their pre-SN properties (e.g., angular momentum distribution) are affected by binary interaction such as mass transfer (e.g., Sana et al. 2012). Applying our calculation to the pre-SN models in such systems deserves a separate work.

Our work has potential application to some new explosive phenomena as well:

\begin{itemize}

\item Super luminous supernova (SLSNe)

SLSNe are 10-100 times more luminous than normal SNe (Gal-Yam 2012), and may be explained by collapsar or magnetar, but the central engine of SLSNe still remains a question (Gal-Yam 2012; Yu et al. 2017).
Recently, Moriya et al. (2018) studied the possibility that the hydrogen-poor SLSNe might be powered by fallback accretion model. They found that the most promising progenitor to explain SLSNe by this model is the one with not too large accreted mass as well as ejecta mass, and the resultant SLSN has a short rise time.
For the pre-SN models we took, the most massive one has a ZAMS mass of $40M_{\odot}$.
We expect that a even more massive star (e.g. ZAMS mass $\gtrsim 50M_{\odot}$) has the potential to produce a SLSN in the disk wind scenario.

\item Fast luminous transients

If only the outermost layer has a sufficient specific angular momentum to form a
disk, the disk will form at a late time.
This might be the case if  the progenitor is a BSG or a WR star, and the outcome could be a fast luminous blue transient (Kashiyama \& Quataert 2015).
KSN2015K (Rest et al. 2018) and AT2018cow (Prentice et al. 2018) could be such examples.
For the pre-SN models we took, Z002M30v37 may produce such a fast luminous transient.

\item Secondary explosion

Following the first disk-wind driven explosion, if a certain amount of envelope material
falls back again, it may result in a secondary, or even multiple, explosion(s) in the same manner.
However, this scenario requires very stringent conditions, such as what Wang et al. (2018) found in explaining the unusual SN iPTF14hls whose light curve has at least five peaks.

\item The disappearing star

The recently discovered `disappearing star' (Adams et al. 2017) is considered to be a failed SNe. Kochanek et al. (2008) gave a detailed discussion about it considering a supergiant progenitor. In our work, there is no disk formation if the specific angular momentum of all envelope is low. This occurs when the metallicity is relatively high so the angular momentum was lost during the star's lifetime via the stellar wind. All the material fall directly into the BH and the star will disappear.
This situation corresponds to the pre-SN models with $Z \gtrsim 10 \%$ of the solar value and $M \gtrsim 40M_{\odot}$.
\end{itemize}

\section*{Acknowledgements}
We are grateful to Rosalba Perna for providing us the series of pre-SN models. We thank the referee for detailed comments and suggestions that greatly helped to improve the manuscript. This work is supported by NSFC Grant No. 11673078.

\end{document}